\documentclass[11pt]{article}
\usepackage{moriond,epsfig}

\def\Journal#1#2#3#4{{#1} {\bf #2}, #3 (#4)}

\def\APJ{\em ApJ}
\def\APJL{\em ApJL}

\def\MNRAS{\em MNRAS}
\def\ASTROPART{\em Astropart. Phys.}

\def\be{\begin{equation}}
\def\ee{\end{equation}}
\def\bea{\begin{eqnarray}}
\def\eea{\end{eqnarray}}

\begin{document}
\vspace*{4cm}
\title{PROSPECTS FOR WEAK LENSING STUDIES WITH NEW RADIO TELESCOPES}

\author{ MICHAEL L. BROWN }

\address{Jodrell Bank Centre for Astrophysics, School of Physics \&
  Astronomy, University of Manchester, Oxford Road, Manchester M13 9PL}

\maketitle\abstracts{ I consider the prospects for performing weak
  lensing studies with the new generation of radio telescopes that are
  coming online now and in the future. I include a description of a
  proposed technique to use polarization observations in radio weak
  lensing analyses which could prove extremely useful for removing a
  contaminating signal from intrinsic alignments. Ultimately, the
  Square Kilometre Array promises to be an exceptional instrument for
  performing weak lensing studies due to the high resolution, large
  area surveys which it will perform. In the nearer term, the e-MERLIN
  instrument in the UK offers the high sensitivity and sub-arcsec
  resolution required to prove weak lensing techniques in the radio
  band. I describe the SuperCLASS survey -- a recently accepted
  e-MERLIN legacy programme which will perform a pioneering radio weak
  lensing analysis of a supercluster of galaxies.}

\section{Weak lensing in the radio band}
Weak gravitational lensing is the coherent distortion in the images of
faint background galaxies due to gravitational light deflection caused
by intervening (dark) matter distributions. On the very largest
scales, the effect traces the large scale structure of the Universe
and is known as cosmic shear. The vast majority of weak lensing
surveys to date have been conducted in the optical bands since large
numbers of background galaxies are required in order to measure the
small ($\sim\!$ a few $\%$) distortions. However, a new generation of
powerful radio facilities is now imminent which makes weak lensing in
the radio band a viable alternative.

The only significant measurement of cosmic shear in the radio band to
date is the work of Chang et al.~(2004) who made a statistical
detection in the Very Large Array (VLA) FIRST survey. Recently a
further attempt to measure radio weak lensing has been applied to data
from the VLA and old MERLIN telescopes (Patel et al.~2010).  This
latter work did not detect a significant lensing signal precisely
because of the small number density of galaxies typically found in
radio surveys. However, it was able to assess the feasibility of doing
so and also proposed that systematic effects could be removed by
observing the same patch of the sky in the radio and optical
wavebands. The e-MERLIN and LOFAR facilities, along with the Square
Kilometre Array (SKA) precursor telescopes, MeerKAT (Karoo Array
Telescope) and ASKAP (Australian Square Kilometre Array Pathfinder),
will be of sufficient sensitivity to achieve a comparable source
galaxy number density to planned optical surveys. Ultimately, all of
these facilities will act as pathfinders for the SKA itself which will
conduct all-sky surveys with unprecedented sensitivity in the radio
band towards the end of this decade.

Performing weak lensing in the radio band is particularly attractive
for a number of reasons. For example, one of the main obstacles facing
the optical lensing community is an issue of instrumental systematics:
an exquisite deconvolution of the telescope point spread function
(PSF) is required in order to return unbiased estimates of the galaxy
shapes. In contrast to complicated and spatially varying optical PSFs,
radio telescopes have highly stable and well understood beam
shapes. In addition, a definitive weak lensing survey conducted with
the SKA would yield precise redshifts for a large fraction of the
source galaxies through the detection of their H1 emission line
(e.g.~Blake et al. 2007). Uncertainties and biases associated with
photometric redshift errors would consequently be greatly reduced with
an SKA lensing survey.

In addition to instrumental systematics, weak lensing surveys are also
subject to serious astrophysical systematics --- \emph{intrinsic
  galaxy alignments}. Galaxies are expected to exhibit some degree of
alignment in their orientations due to the tidal influence of
large-scale structure during the galaxy formation process. These
intrinsic alignments can mimic a cosmic shear signal and represent one
of the biggest challenges for precision cosmology measurements using
weak lensing.

With this in mind, a unique advantage offered by measuring lensing in
the radio band is the polarization information which is usually
measured in addition to the total intensity in radio surveys. Previous
authors have exploited the fact that the polarization position angle
is unaffected by lensing in order to measure gravitational lensing of
distant quasars (Kronberg et al. 1991, 1996; Burns et al. 2004). In a
recent paper with R. Battye (Brown \& Battye 2011a), I showed how one could extend
this idea to measure cosmic shear. The technique relies on there
existing a reasonably tight relationship between the orientation of
the integrated polarized emission and the intrinsic morphological
orientation of the galaxy. The existence of this relationship needs to
be established for the high-redshift star-forming galaxies which are
expected to dominate the radio sky at the $\mu$Jy flux sensitivities
achievable with forthcoming instruments. However, such a relationship
certainly exists in the local universe (Stil et al. 2009) and it is
reasonable to assume that it persists to higher redshift.  A key
difference between the polarization technique and standard techniques
for measuring lensing is that the former does not assume that the
ensemble average of the intrinsic shapes of galaxies vanishes. It is
thus, in principle, able to cleanly discriminate between a lensing
signal and a possible contaminating signal due to intrinsic galaxy
alignments (Brown \& Battye 2011a).

\begin{figure}
\psfig{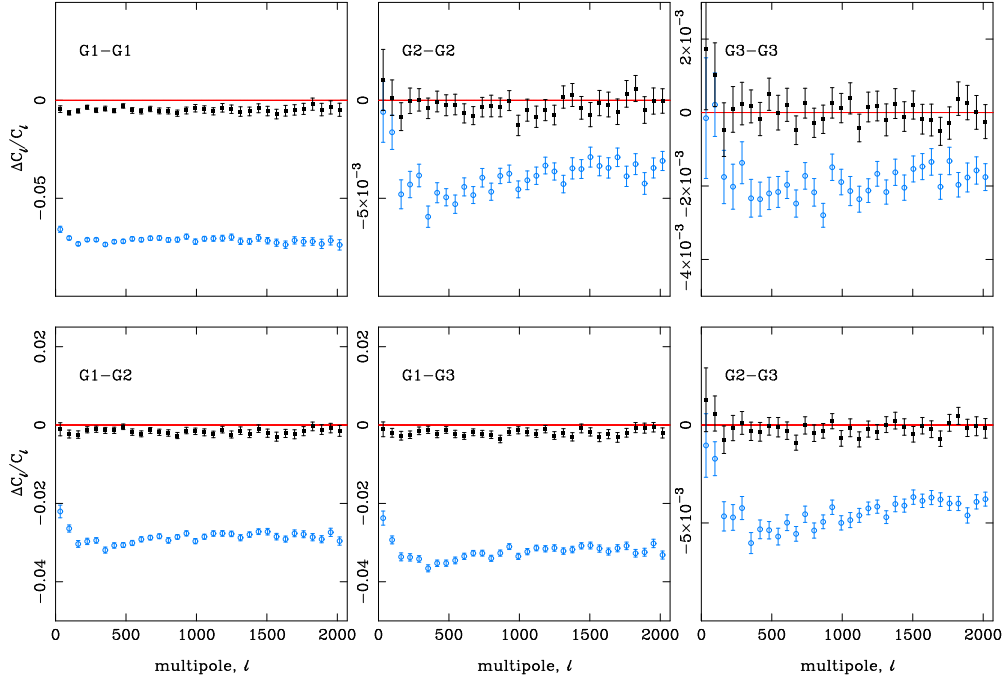}
\caption{Fractional bias ($\Delta C_\ell / C_\ell$) in the simulated
  reconstruction of the shear auto- and cross-power spectra in three
  redshift bins in the presence of a contaminating signal from
  intrinsic alignments. The top panels show the residuals in the
  auto-power spectra in the three redshift bins (increasing in
  redshift from left to right). The bottom panels show the cross-power
  residuals. The light blue points show the result obtained using a
  standard lensing analysis and shows a clear bias due to the
  intrinsic alignment effect. The black points show the recovery
  obtained with an analysis using polarization information where the
  bias is reduced by an order of magniutde.
\label{fig:fig1}}
\end{figure}

Fig.~1 demonstrates the potential power of this technique in terms of
its ability to mitigate intrinsic alignments. The figure shows the
bias in the recovered weak lensing power spectra in simulations of a
future SKA-like survey in the presence of a contaminating signal from
intrinsic alignments. For these simulations, we assumed that the
orientation of the polarized emission is an unbiased tracer of the
intrinsic structural position angle with a scatter of 5 degs and that
we can measure the polarization in 10\% of the total galaxy sample.

\section{The SuperCLASS survey}\label{superclass}

e-MERLIN, the UK's next generation radio telescope, has now been
commissioned and has recently begun science operations. It consists of
seven radio telescopes, spanning 217 km, connected by a new optical
fibre network to Jodrell Bank Observatory near Macclesfield in the
UK. Of the present (or soon to be available) radio instruments
e-MERLIN has a number of advantages for detecting weak lensing in the
radio band. Most significant of these is the fact that it has very
high resolution ($\approx 0.2$ arcsec at L-band) making it possible to
detect the ellipticity of individual sources since they have similar
angular extent to that detected in the optical (Muxlow et al.,
2005). The SuperCLuster Assisted Shear Survey (SuperCLASS; P.I. R.
Battye) was recently approved as an e-MERLIN legacy project to pursue
the objective of performing weak lensing analyses in the radio
band. SuperCLASS will survey a 1.75 degs$^2$ region of sky with 0.2
arcsec at 1.4 GHz to an unprecedented r.m.s.~sensitivity level of 4
$\mu$Jy bm$^{-1}$. In addition to performing a standard weak lensing
analysis, these data will allow us to perform the first demonstration
of the polarization lensing techniques described above on real data.

The presently chosen target is a region containing 5 Abell clusters at
right ascension $\approx$ 14 hours and declination $\approx$ 68 degs
with measured redshifts $\approx$ 0.2. All five clusters (A968, A981,
A998, A1005, A1006) have been detected by ROSAT with luminosities
compatible with them having masses in the range $(1 - 2) \times
10^{14}$ M$_\odot$.  We expect to be able to detect the weak lensing
effect of these clusters and also from some of the large-scale
filamentary structure expected to permeate the regions between the
clusters.

Fig.~2 shows a simulation of how well we might expect to recover the
dark matter distribution in the region of the supercluster. It shows
reconstructions of the dark matter distribution in a randomly chosen
1.75 deg$^2$ region of simulated sky as seen in the N-body
simulations of White (2005). The projected mass reconstructions were
performed using the algorithms described in Brown \& Battye (2011b)
which extended standard mass-reconstruction techniques to include
potential information coming from polarization observations. The
reconstructions are presented for sensitivity levels approximating the
SuperCLASS survey and for a sensitivity level approximating what one
might expect to achieve with the SKA.

\begin{figure}
\psfig{figure=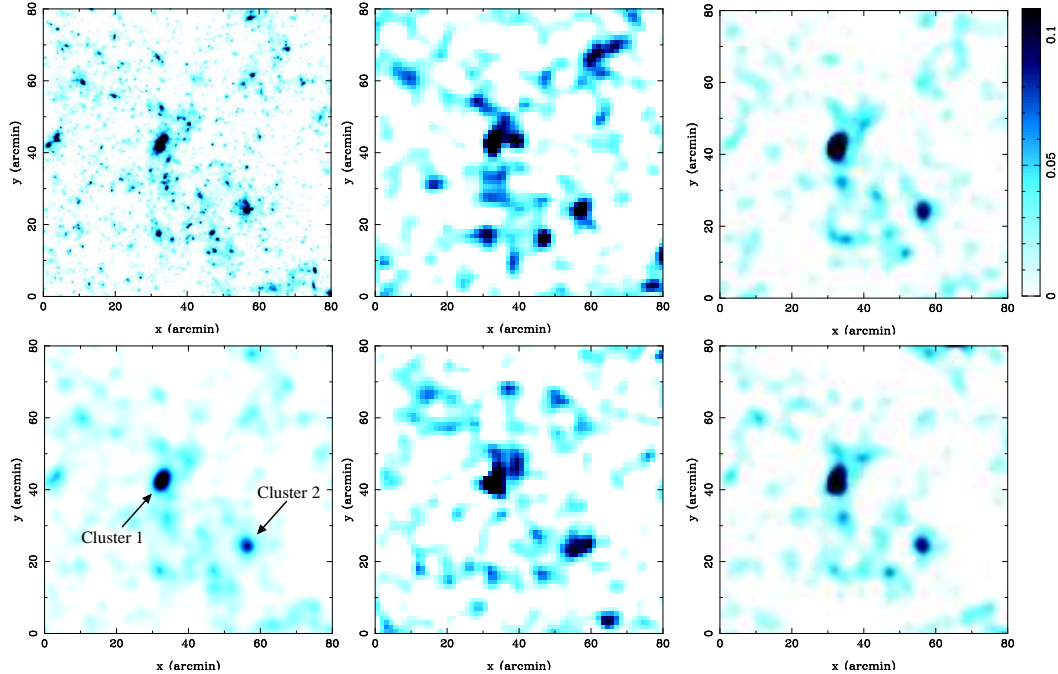,height=3.5in,angle=0}
\caption{Simulation of the recovery of the dark matter distribution in
  a randomly selected simulation designed to mimic the SuperCLASS
  survey and a future SKA survey. The input distribution is shown in
  the top-left panel and shown smoothed in the bottom left. The middle
  panels show the recovery for a SuperCLASS-like survey with e-MERLIN
  and the right hand panels show the simulated recovery for future
  surveys with the SKA. The top panels show the recoveries obtained
  using a standard lensing analysis. The bottom panels show the
  recovery obtained using the polarization technique.  
\label{fig:radish}}
\end{figure}

\section{Conclusion}\label{conclusion}

I have given a brief summary of the status of the field of weak
lensing in the radio band. While it currently lags well behind the
field of optical weak lensing, the new radio instruments coming online
now make radio weak lensing a viable alternative which is
complementary to large scale optical surveys. In particular, radio
polarization observations offer interesting possibilities for removing
intrinsic alignments from radio lensing surveys. Over the course of
the next few years, the SuperCLASS survey on the e-MERLIN telescope
will act as a pathfinder experiment for more ambitious radio lensing
surveys with future instruments. 

\section*{Acknowledgments}
I thank the SuperCLASS collaboration, and in particular, the P.I. and
my close collaborator Richard Battye, for their contributions to this
work. I thank the STFC for an Advanced Fellowship (reference
ST/I005129/1) and the ERC for the award of a Starting Independent
Researcher Grant (No. 280127).

\end{document}